\documentclass[preprint]{aastex}
\usepackage{amsmath,amssymb,graphicx}

\newcommand{\tmop}[1]{\ensuremath{\operatorname{#1}}}

\newcommand{\tmtextit}[1]{{\itshape{#1}}}

\shorttitle{Discovery of a young supernova remnant G354.4+0.0}
\shortauthors{Roy et al.}

\begin{document}

\title{Discovery of a small diameter young supernova remnant G354.4+0.0}
\author{Subhashis Roy}
\affil{NCRA-TIFR, Pune-411007, India}
\email{roy@ncra.tifr.res.in}

\author{Sabyasachi Pal}
\affil{Indian Centre for Space Physics, Kolkata-700084, India}
\affil{Ionospheric and Earthquake Research Centre, Sitapur-7211154, India}
\email{sabya@csp.res.in}

\begin{abstract}

We report discovery of a shell like structure G354.4+0.0 of size 1.6$'$ that
shows morphology of a shell supernova remnant.
Part of the structure show polarized emission in NRAO VLA sky survey (NVSS)
map. Based on 330 MHz, 1.4 GHz Giant Metrewave Radio
Telescope (GMRT) observations and existing observations at higher frequencies,
we conclude the partial shell structure showing synchrotron emission is
embedded in an extended HII region of size $\sim 4'$.
The spectrum of the diffuse HII region turns over between 1.4 GHz and 330
MHz. HI absorption spectrum shows it to be located more than 5 kpc away from
Sun. Based on morphology, non-thermal polarized emission and size, this object
is one of the youngest supernova remnants discovered in the Galaxy with 
an estimated age of $\sim 100-500$ years.

\end{abstract}

\keywords{supernova remnants -- radio lines: ISM -- H II regions -- radio
continuum: general}

\section{Introduction}

Galactic OB star counts, pulsar birth rates, iron abundance and estimates of
supernova rates in Local Group galaxies suggest that there should be more than
1000 supernova remnants (SNRs) in our Galaxy \citep{TAMMANN1994}. However, only
274 are presently cataloged \citep{GREEN2009}. This lack of detection is
partly due to poor angular resolution of existing surveys which disfavors
detection of small diameter SNRs \citep{GREEN1991} that future surveys with
upcoming facilities \citep{KASSIM2005,BOWMAN2013} could address.
Supernova rate in our Galaxy is about 2.8 per century \citep{LI2011}. Hence, in
the last 400 years more than ten supernova explosions have taken place in the
Galaxy. However, only two of them (Cas A and G1.9+0.3) have been reported.
There are only three SNRs of angular size $\lesssim$1.5$'$ listed in the
catalog of \citet{GREEN2009}. They are G1.9+0.3, G54.1+0.3 and G337.0$-$0.1.
Among them, G1.9+0.3 has been shown to be the youngest known SNR
\citep{REYNOLDS2008} in the Galaxy with age of about 150 years. The SNR
G54.1+0.3 is centrally powered by a pulsar and the size of the emission region
is 1.5$'$.  The actual size of the outer shell is unknown \citep{LEAHY2008} and
hence the remnant age could be much more than a few hundred years. The
size of the SNR G337.0$-$0.1 has been measured to be 5 pc \citep{SHARMA1997},
that suggests it to be a candidate young SNR. 

The line of sight towards the Galactic center passes the longest distance
through the Galaxy, that enhances the probability of detecting an SNR.
Therefore, we had conducted observations of a few regions close to the Galactic
center (GC) with the Giant Metrewave Radio Telescope (GMRT) at 330 MHz
\citep{ROY2006}. The main objective of these observations was to confirm the
nature of certain candidate SNRs.  From one of these fields, a small diameter
($\sim$1.5$'$) object G354.46+0.07 with partial shell morphology was
identified. To confirm its nature, this field was re-observed with the GMRT
at 1.4 GHz and 330 MHz.  Here we describe its confirmation as a newly
discovered young SNR in the Galaxy. In Sect.~2, observations and data analysis
procedures are described.  Results and Discussions are presented in Sect. 3
and 4 respectively.

\section{Observations and data analysis:}

The field mentioned above was observed on 17th and 18th June 2007 at 1.4 GHz
and 330 MHz respectively. In the absence of system temperature measurement at
GMRT, we have corrected for the variation of sky temperature from calibrator
to the target source following \citet{ROY2004} keeping the automatic level
control of the antennas off
\footnote {NCRA technical report at http://ncralib1.ncra.tifr.res.in:8080/jspui/handle/2301/437}.
During the 330 MHz and 1.4 GHz observations, each of the 2 IFs had a bandwidth
of 4 MHz and 16 MHz respectively. Both the IFs during the above set of
observations had 128 spectral channels. At 1.4 GHz, it provided a velocity
resolution of 6.6 km~s$^{- 1}$ and a velocity coverage of more than $\pm$400
km~s$^{- 1}$. 3C48 was used as primary flux density calibrator in both the
frequency bands. HI absorption towards 3C287 is known to be less than 1\%
\citep{DICKEY1978} and was used as bandpass calibrator at 1.4 GHz. 1751$-$253
and 1714$-$252
\footnote{VLA calibrator catalog from http://www.vla.nrao.edu/astro/calib/manual/csource.html} were used
as phase calibrators at 1.4 GHz and 330 MHz respectively. The 1.4 GHz data was
analyzed following the standard procedures in AIPS
\footnote{http://www.aips.nrao.edu}. After preliminary editing, time and
frequency based calibration, a continuum image of the object was made leaving
aside the channels showing evidence of strong absorption. This image was used
to phase only self calibrate the data for all the frequency channels.  Before
making the channel maps, the AIPS task UVLSF was used to subtract a constant
term with a constant slope across the frequency channels corresponding to the
continuum from the visibility data. To reduce the effect of comparatively
large-scale structures in atomic hydrogen (HI) along the line of sight, we have
done a high-pass filtering with a short\tmtextit{ uv} cut-off of 1 k$\lambda$
that resolves out structures with sizes $>$3$'$ while making the spectral line
maps for detection of HI absorption.

During analysis of the 330 MHz data, after calibration and editing, each 10
adjacent frequency channels were averaged providing a channel width of 1.25
MHz in the output data. This process significantly reduces data volume while
keeping bandwidth smearing smaller than the synthesized beam during imaging up
to half power point of the antennas.  The initial images were improved by
phase-only and later by amplitude \& phase self-calibration.  The final image
is made using multi-resolution CLEAN \citep{WAKKER1988}.

\section{Results:}

\subsection{Continuum images}

The continuum map of this source made from 330 MHz data having a resolution of
26$^{''} \times 11^{''}$ is shown in Fig.~1. This map is sensitive to large
scale structures of size up to 30$'$. The size of the object G354.4+0.0 is
$\sim$1.5$'$ and shows morphology of a partial shell. To detect any polarized
emission and thereby confirm non-thermal emission, we have searched the NRAO
VLA sky survey (NVSS) \citep{CONDON1998} map at the same location of the sky, and
the polarization vectors from NVSS (resolution 45$^{''}$) are overlaid on Fig.
1. To make the polarization total intensity and angle images from the NVSS
Stokes Q and U maps (rms noise $\sim$0.3 mJy~beam$^{- 1}$), all pixels below a
signal to noise ratio of 4.5 were blanked and correction for noise bias in
polarized total intensity was made using the AIPS task COMB (option
POLC). We do detect significant polarized emission from near the two brightest
parts of the shell like structure with peak polarized flux densities of
$\sim2.2 \pm$ 0.3 mJy~beam$^{- 1}$ (Fig.~1). The percentage polarization of the
northern and southern components are 0.6$\pm0.1$ percent and 0.75$\pm0.1$ percent
respectively.  

The 1.4 GHz continuum map of the shell without any short \tmtextit{uv} cut-off
showed extended emission (size $\sim 4'$). To
image the shell structure at a higher resolution and to measure its flux
density, we resolved out any extended structure of size $> 3'$ using a short
\tmtextit{uv} cut-off of 1000 $\lambda$ to the data during imaging. We also
avoided the frequency channels having HI absorption during imaging. The
resultant image is shown in Fig.~2 (left).
As can be seen from Fig.~1 and 2 (left), the morphology of the source is of shell
type. To measure the size of this object, we took four cross-cuts across the shell at
different orientations passing through the center of the shell like structure
at 330 MHz.  From the angular separation of the two peaks in the cross-cuts, we
measured the angular size of the shell. The mean angular diameter of the shell
is found to be $\sim94^{''}$ or about 1.6$'$. 
Rms error on the mean angular diameter is $\sim 5^{''}$.  To measure flux
density at 330 MHz and spectral index of the
shell between 330 MHz and 1.4 GHz, this object was further imaged from 330 MHz
data with a short \tmtextit{uv} cut-off of 1000 $\lambda$ (same as at 1.4
GHz). This minimizes flux density variation due to different sensitivity in
imaging large scale structures at 330 MHz with respect to the 1.4 GHz
map. To compare flux densities in these two frequencies, the 1.4 GHz map was
convolved to the resolution of the new 330 MHz map and brought to the same
pixel size and center. 
AIPS task BLSUM was used to measure flux densities covering the same polygon
region in both the maps.  Measured flux densities are 0.7$\pm$0.1 and
0.9$\pm$0.1 Jy at 1.4 GHz and 330 MHz respectively.  The error quoted in the
flux density is 1$\sigma$ and is estimated from map rms and uncertainty in
absolute flux density calibration ($\sim 5$\%).  The observed spectral index
between 1.4 GHz and 330 MHz is $-$0.2$\pm$0.1, that is quite flat and
unexpected for a shell type supernova remnant. We note that the shell show
extension towards the East at 330 MHz (e.g., Fig.~1) that is not seen in
the 1.4 GHz image in Fig.~2 (left). The 330 MHz flux density of the shell when
this emission is included is 1.1$\pm$0.1 Jy.

\subsection{Nature of diffuse emission around the shell}

To image the diffuse emission, we made a low resolution continuum map shown in
Fig.~2 (right) of the field at 1.4 GHz after subtracting the significant CLEAN
components ($>2 \sigma$) of the map shown in Fig.~2 (left) from the
\tmtextit{uv} data. 
An extended source of size $\sim$4$'$ is seen in Fig.~2 (right) coincident with
the shell G354.4+0.0. This extended source has also been cataloged in
the PMN survey at 4.8 GHz \citep{WRIGHT1996} as J1731$-$3334. We have also
detected the complex at 843 MHz from the Molonglo Galactic plane survey
\citep{GREEN1999}. The measured flux densities of the diffuse emission and the
shell at various frequencies and their physical parameters are given in
Table~1.  Note that the flux densities of the shell and the diffuse emission
at 330 MHz given in Table~1 were measured from the large scale sensitive map
shown in Fig.~1.

The flux density of the diffuse emission decreases significantly at 330 MHz
as compared to 1.4 GHz and this indicates self absorption.
For a typical extended source in the Galaxy, self absorption at meter
wavelengths is caused by bremsstrahlung process. No significant free-free
absorption by the line of sight gas is likely at 330 MHz as significant optical
depth of this gas is observed only at $\lesssim$100 MHz
\citep{BROGAN2003}. The free-free optical depth ($\tau$) is related to
observing frequency ($\nu$) by 
\begin{equation}
\tau= \int^L 0.2 n^2_e T^{- 1.35}_e \nu^{- 2.1} \tmop{dL}.  
\end{equation}
Where, $n_e$ is electron density, $T_e$ electron temperature and
$L$ is the path length through the diffuse HII region. 

The flux density (S) of an object is related to its brightness temperature in
radio frequency ranges by
\begin{equation} 
   S = \frac{2 \tmop{kT}_b \nu^2 \Delta \Omega}{c^2}
\end{equation} 
Where, $\Delta \Omega$ is the solid angle and $T_b$ is the brightness
temperature of the diffuse emission. In case of bremsstrahlung emission, the
brightness temperature of the emission is related to the thermal electron
temperature ($T_e$) by 
\begin{equation}
T_b = T_e \times (1- e^{-\tau}). 
\end{equation}
We note that significant decrease of flux density of the diffuse emission at
330 MHz (Table~1) indicates significant optical depth at this frequency.
However, the measured value of $T_b$ for the HII region at 330 MHz
(using Table~1) is only about 500 K. As this temperature would be
comparable to the Galactic background temperature at 330 MHz, correction due to
absorption of the Galactic background radiation by the HII region needs to be
made. We measured the Galactic background temperature from \citet{HASLAM1982}
408 MHz single dish map convolved to the primary beam of the GMRT at 330 MHz
and scaled to the same frequency assuming a spectral index of $-$2.7 for the
sky brightness temperature.  The measured temperature is 700 K towards the 
HII region. One sigma error on this measurement is less than 70 K. 
Following the discussion in \citet{NORD2006}, we can write\\
$T_b = T_e \times (1- e^{-\tau}) +T_{gb}e^{-\tau} +T_{gf}$, \\
where $T_{gb}$ and $T_{gt}$ are the brightness temperature of the non-thermal
emission behind and in front of the diffuse HII region respectively.
For a direction slightly away from the HII region, the Galactic plane
temperature ($T_{gt}$) is \\ 
$T_{gt}= T_{gb} + T_{gf}$.\\
An Interferometer is insensitive to the total large scale temperature (or flux
density) and measures the difference between the above two, giving an observed
brightness temperature of the HII region of 
\begin{equation}
T_{obs}= T_e \times (1- e^{-\tau}) - T_{gb} \times (1- e^{-\tau}).\\
\end{equation}
Therefore, $T_b$ in Eq.~2 needs to be replaced by $T_{obs}$ (Eq.~4), where
$\tau$ is given by Eq.~1. One finally gets,
\begin{equation}
   S = 3.07\times 10^8 \times \nu^2 \times [T_{e4}-(T_{gb4}\times \nu^{-2.7})] \times [1 -e^{-0.306 \times EM \times (T_{e4}^{-1.35}) \times \nu^{-2.1}}] \times \Delta \Omega
\end{equation}

where, $T_{e4}$ $-$ electron temperature in 10$^4$~K, $T_{gt4}$ $-$ Galactic
plane temperature at 1 GHz in 10$^4$~K, EM $-$ emission measure in the unit of
$10^6$~cm$^{-6}$~pc and $\nu$ is in GHz. Eq.~5 is fitted to the observed flux
densities of the diffuse HII region to measure its physical properties ($T_e$,
$n_e$).

At 843 MHz and 4.8 GHz, the contribution of the shell could not be separated
from the diffuse emission. Therefore, estimation of flux densities of the
diffuse emission at these two frequencies depend on the actual flux density
of the shell emission that in turn depends on its location with respect to the
diffuse HII region. Therefore, we consider below three possible configuration
of the shell with respect to the HII region, thereby allowing us to constrain
the spectral index of the shell.

\subsubsection{G354.4+0.0 shell at the center of the diffuse HII region}

From the plot of the flux densities of the shell and the diffuse emission
(Fig.~3), we note the slopes of their flux densities between 1.4 GHz and 330
MHz are very different, indicating the shell type object and the diffuse
emitting region are two different objects.  When the shell is at the center of
the HII region, its emission undergoes absorption by the diffuse HII region at
330 MHz and the intrinsic spectrum of the shell is steeper than the observed
one as given in Table~1.  By assuming the diffuse emission to be optically thin
($\tau << 1$) at 843 MHz and above, and an intrinsic spectral index of $-$0.5
for the shell (typical for a shell type SNR \citep{GREEN2009}), we estimate
its flux densities at 843 MHz and 4.8 GHz from its measured flux density at
1.4 GHz. The estimated flux densities are then subtracted from the total
measured flux densities of the complex at the above two frequencies to get the
flux densities of the diffuse emission. 

If we know the distance to the HII region, $T_{gb}$ could be estimated from our
measured temperature of $T_{gt}=$~700~K towards G354.4+0.0. Here, we assume the
shell structure and the diffuse HII region is located nearly at the distance of
the Galactic center, but is at the near side of the GC (discussed in detail in
Sect. 3.3).  Due to expected high density of massive stars in the GC region, it
is likely that most of the Galactic non-thermal emission is generated in the
region.  Therefore, we assume $T_{gb}= T_{gt}$, and scale the value of $T_{gt}$
as a function of frequency ($T_{gt}=35 \times \nu^{-2.7}$). The flux densities
of the diffuse emission as a function of frequency 
fitted well by Eq.~5 (see Fig.~3), with a reduced Chi-square of 0.7. From
the fitted parameters, emission measure and electron temperature of the
HII region are 1.7$\pm0.09 \times 10^4$ cm$^{-6}$~pc and 2300$\pm$350~K
respectively.  Assuming the source to be spherical and located at the distance
of the Galactic center (8 kpc), path length through the source ($4'$) is about
9 pc.  Consequently, the electron density in the diffuse emitting region is
estimated to be 44$\pm$1 cm$^{-3}$. This gas will have a $\tau$ of about 0.4 at
330 MHz.

Given the relative location of the shell with respect to the diffuse emission,
the emission from the shell would undergo free-free absorption with $\tau$
being half of that of the ionized gas and $\sim 20$ percent of its emission
would be absorbed at 330 MHz. Moreover, its brightness temperature and
consequent flux density is underestimated in Table~1 due to high large scale
diffuse Galactic background (Eq.~4). Given the angular cross-section used (4.4
square arc-min) to measure its flux density (Table~1) and $T_{gb}$ of 700 K at
330 MHz, the flux density that was underestimated in Table~1 at 330 MHz is 0.28
Jy. Once this flux density is added with the observed flux density of the shell
at 330 MHz, and correction is made for the fraction that was absorbed by the
diffuse HII region, its estimated flux density at 330 MHz is 1.5 Jy. This shows
its spectral index between 1.4 GHz and 330 MHz is about $-$0.5, consistent with
spectral indices of typical shell type SNRs in the Galaxy \citep{GREEN2009}.

\subsubsection{G354.4+0.0 shell in front of the diffuse HII region}

In this case, the shell's emission do not undergo any absorption by the HII
region. Considering its measured flux densities at 1.4 GHz and 330 MHz
(Table~1) and underestimation of its flux density due to large scale Galactic
plane emission as discussed above, its spectral index is $-$0.36. By using its
flux density at 1.4 GHz and the spectral index, we estimate its flux densities
at 4.8 GHz and 843 MHz, which are then subtracted from the corresponding total flux
densities of the complex to get the flux densities
of the diffuse HII region at the two frequencies (Table~1). A fit of Eq.~5 to
the new set of flux densities showed the emission measure and electron
temperature to remain the same as before within the error bars.

\subsubsection{G354.4+0.0 shell behind the diffuse HII region}

If the SNR is located behind the HII region, the absorption of the shell
emission would be more than the case when it was assumed to be at the center
of the HII region. At 843 MHz and 4.8 GHz, only the total flux densities of
the complex are available, and the flux densities of the diffuse HII region
were computed by subtracting the flux density of the shell with an assumed
spectral index of $-$0.5. In the present case, we assume that the properties
of the diffuse HII region do not change significantly with changes in flux
densities of the shell (quite small with respect to the HII region) at 843 MHz
and 4.8 GHz. 
Optical depth towards the shell at 330 MHz is about 0.4 (Sect.~3.2.1). Based on
its 1.4 GHz flux density and underestimation of its flux density at 330 MHz due
to large scale Galactic plane emission as discussed earlier, its spectral index
is estimated to be $-$0.6.  By using its flux density at 1.4 GHz and the
spectral index, we estimate its flux densities at 4.8 GHz and 843 MHz, which
are then subtracted from the corresponding total flux densities of the complex
to get the flux densities of the diffuse HII region at the two frequencies
(Table~1). A fit of Eq.~5 to the set of flux densities showed the emission
measure and electron temperature do remain the same as in Sect~3.2.1 within the
error.

\begin{table}[!htb]
   \caption{Measured parameters of G354.4+0.0 shell and the diffuse HII region}
   \small
  \begin{tabular}{|c|c|c|c|c|c|c|c|c|l|}
\hline
Source & Angular & S$_{330}$ & S$_{843}$ & S$_{1400}$ & S$_{4800}$ & $\alpha_{1000}$ & $\tau_{330}$ & T$_e$ & n$_e$ \\
name   & size    &        &              &            &            &             &        &         &  \\
            & ($'$)   & (Jy)      & (Jy)      & (Jy)      & (Jy) &         &  & (K) & (cm$^{-3}$) \\
\hline
G354.4+0.0 & 1.6$\pm$.1 & 0.9$^a\pm$0.1 & 0.9$^b$      & 0.7$\pm$0.1 & 0.4$^b$  &   $-0.2\pm0.1$  &  $-$ & $-$  & $-$  \\
shell      &            &               &              &             &          &                 &      &      &  \\
\hline
Diffuse & $\sim$4   & 2.4$\pm$0.2 & 3.3$^b\pm$0.3 & 3.5$\pm$0.3 & 3.3$^b\pm$0.2 &  $-0.1\pm0.1$ &  0.4 &  2300 & 44$\pm$1  \\
HII region &        &             & 3.5$^c\pm$0.3 &           & 3.2$^c\pm$0.2  &               &      &     $\pm$350 &  \\
           &        &             & 3.2$^d\pm$0.3 &           & 3.4$^d\pm$0.2  &               &      &              &  \\
\hline

\multicolumn{10}{l}{In the table, S$_{330}$, S$_{843}$, S$_{1400}$ and S$_{4800}$ denote flux densities at 330, 843, 1400 and 4800 MHz} \\
\multicolumn{10}{l}{respectively. $\alpha_{1000}$ $-$ observed spectral index between 1.4 GHz and 843 MHz (HII region) or 1.4 GHz } \\
\multicolumn{10}{l}{and 330 MHz (shell). $\tau_{330} -$ optical depth at 330 MHz.} \\
\multicolumn{10}{l}{$^a$ $-$ 1.1$\pm$0.1 Jy with emission from Eastern part of
shell.} \\
\multicolumn{10}{l}{$^b$ estimated assuming the shell is at the center of the diffuse HII region.}\\
\multicolumn{10}{l}{$^c$ estimated assuming G354.4+0.0 shell is in front of the diffuse emission.} \\
\multicolumn{10}{l}{$^d$ estimated assuming G354.4+0.0 shell is behind the diffuse emission.} \\
  \end{tabular}
  \normalsize
\end{table}

\begin{figure}[htb]
   \centering
  \begin{minipage}[t]{0.45\textwidth}
  \includegraphics[width=\textwidth, clip=true,]{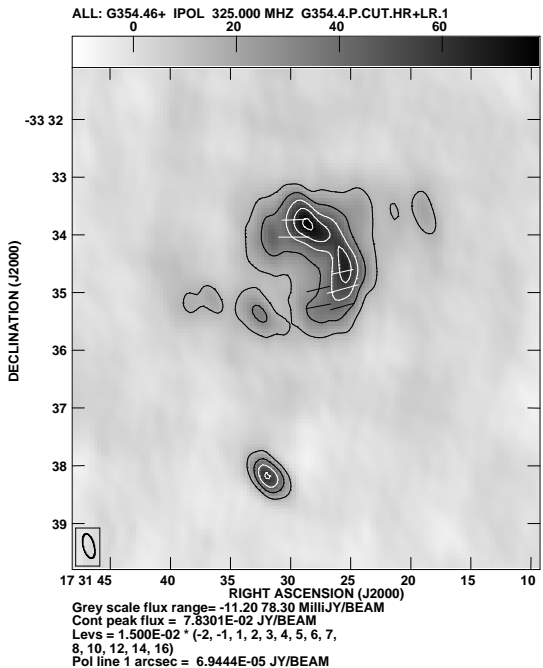}
  \caption{330 MHz continuum map (Gray scale and contour) of the source
  G354.4+0.0 with a resolution of 26$^{''} \times 11^{''}$ with an rms noise of
  3 mJy~beam$^{- 1}$.  Superimposed are the polarization vectors from 1.4
  GHz NVSS map (resolution 45$^{''}$). A polarized flux density of 1.0 mJy
  corresponds to 15$^{''}$ of the polarization vectors.} 
  \end{minipage}
\end{figure}
\begin{figure}[htb]
   \centering
\begin{minipage}[t]{0.9\textwidth}
   \includegraphics[width=\textwidth, clip=true,]{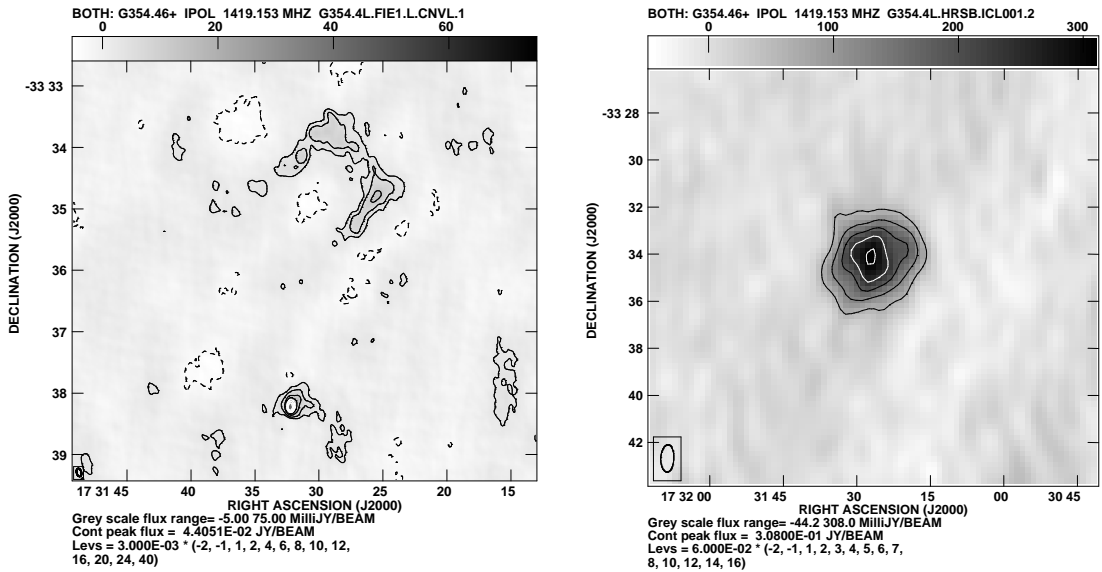}
  \caption{1.4 GHz high resolution continuum map of the source G354.4+0.0
  (left). The resolution is 8$'' \times 4'' .4$, and rms is 1.5
  mJy~beam$^{-1}$.  In right, 1.4 GHz low resolution map of the same field
  after subtracting the source in left. The resolution is 70$^{''} \times
  33^{''}$ and the rms noise is 10 mJy~beam$^{- 1}$.}
  \end{minipage}
\end{figure}

\begin{figure}[htb]
   \centering
  \begin{minipage}[t]{0.45\textwidth}
  \includegraphics[width=\textwidth, clip=true,angle=270,totalheight=6cm]{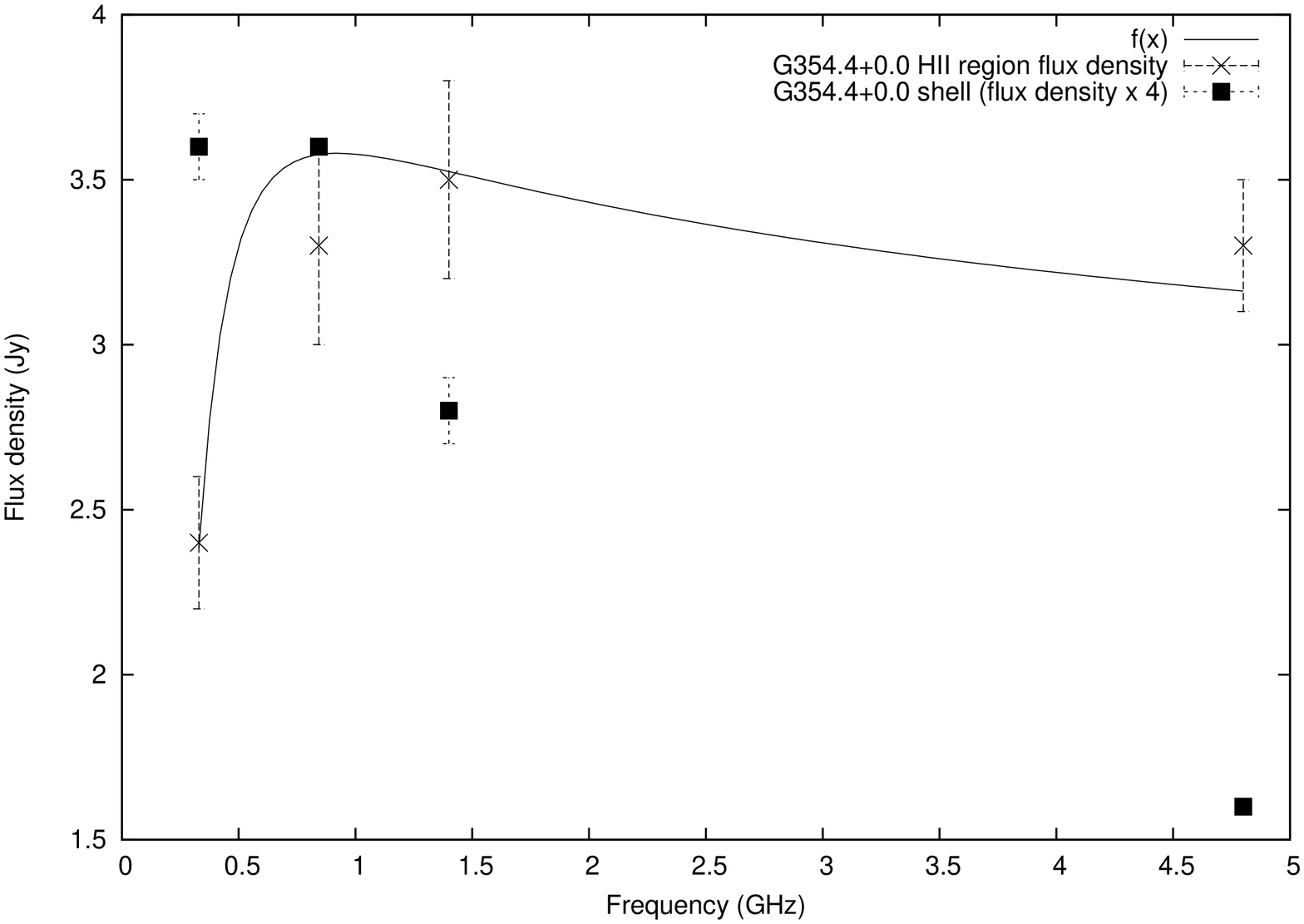}
  \caption{Fit of Eq.~5 (f(x)) to the flux densities of the diffuse HII region
  at different frequencies (cross signs) assuming the shell at the centre
  of the diffuse emission.  Also plotted are the flux densities for the
  G354.4+0.0 shell (with solid squares) scaled by a factor of 4. Its flux
  densities at 843 MHz and 4.8 GHz are only estimates and are shown with no
  error bar.}
  \end{minipage}
\end{figure}


\subsection{Distance from HI absorption}

To measure the distance to the shell type object G354.4+0.0, we plot in Fig.~4
the HI spectrum towards this object. Strong absorption is seen near 0 km~s$^{-
1}$ that is believed to be caused by local gas rotating almost
perpendicular to our line of sight seen near the direction of the GC. HI
absorptions are also seen near $-$40 and $-$80 km~s$^{- 1}$. The line of
sight velocity of the 3 kpc arm at the longitude of this SNR is about $-$80
km~s$^{- 1}$ \citep{COHEN1976}. Presence of absorption by this feature
shows the distance of G354.4+0.0 to be at least 5 kpc from Sun. We also
examined HI absorption spectrum of a few sources in the field and the spectrum
of the brightest small diameter source (peak flux density 0.49 Jy~beam$^{-
1}$) with the RA=17h31m20.7s, DEC=$-$33$^{\circ}$53$'$25$^{''}$ (J2000) is
shown in Fig.~5 (henceforth will be called source A). Compared to Fig.~4, it
shows a much wider absorption profile between +15 to $-$40 km~s$^{- 1}$,
suggesting source A to be much further away than G354.4+0.0.
Flux density of source A is found to go down by more than a factor of four
between 1.4 GHz and 330 MHz indicating it to be a compact HII region. It is
likely that the source A is located at the other side of the Galaxy. Based on
the above, it is likely that G354.4+0.0 is located between 5 kpc to the
Galactic center distance of 8 kpc.

\begin{figure}[htb]
   \centering
\begin{minipage}[t]{0.45\textwidth}
  \includegraphics[width=\textwidth, clip=true,]{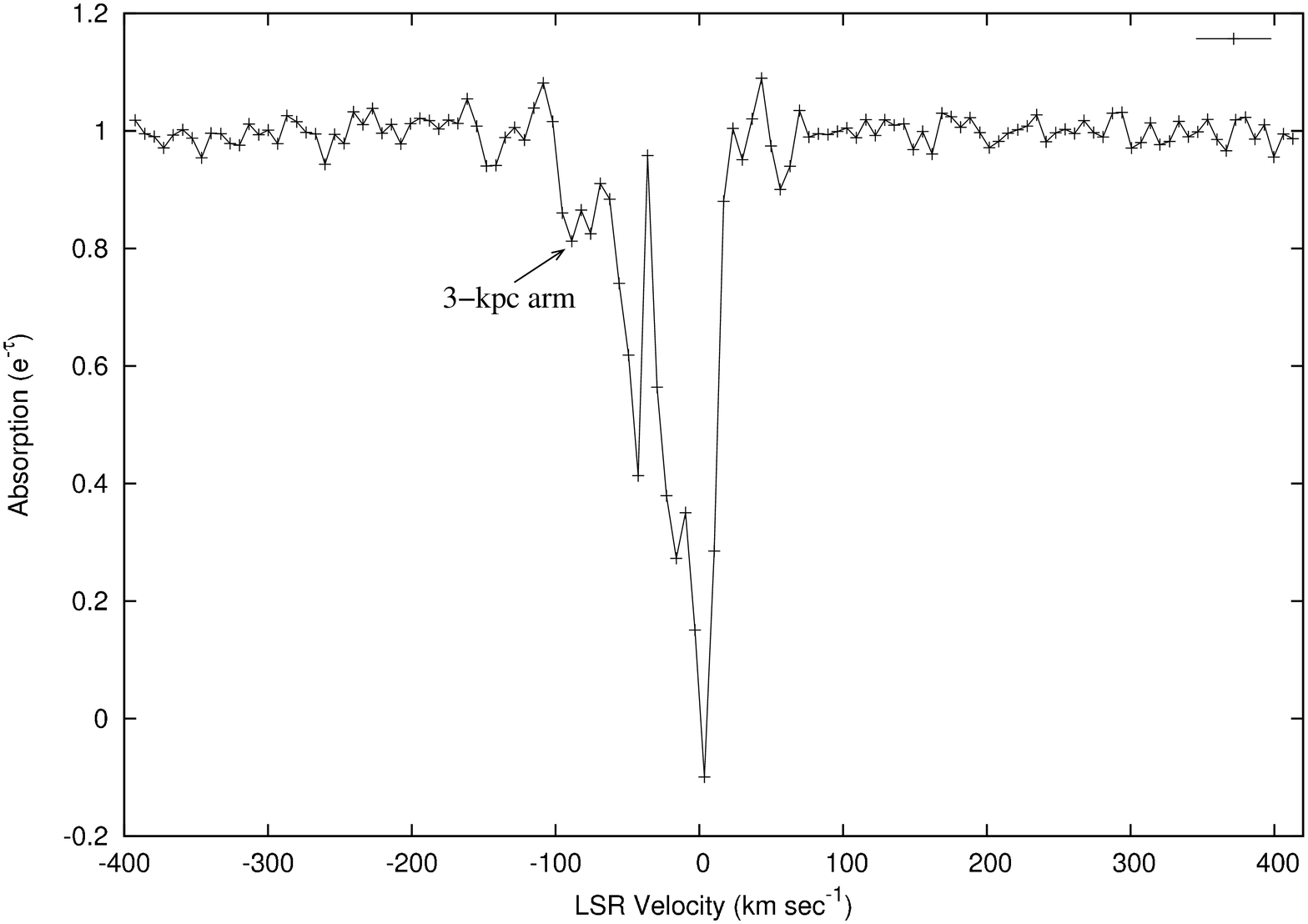}
  \caption{HI absorption spectrum towards G354.4+0.0. Velocity resolution 6.7
  km~s$^{-1}$.}
\end{minipage}
\hfil
\begin{minipage}[t]{0.45\textwidth}
  \includegraphics[width=\textwidth, clip=true,]{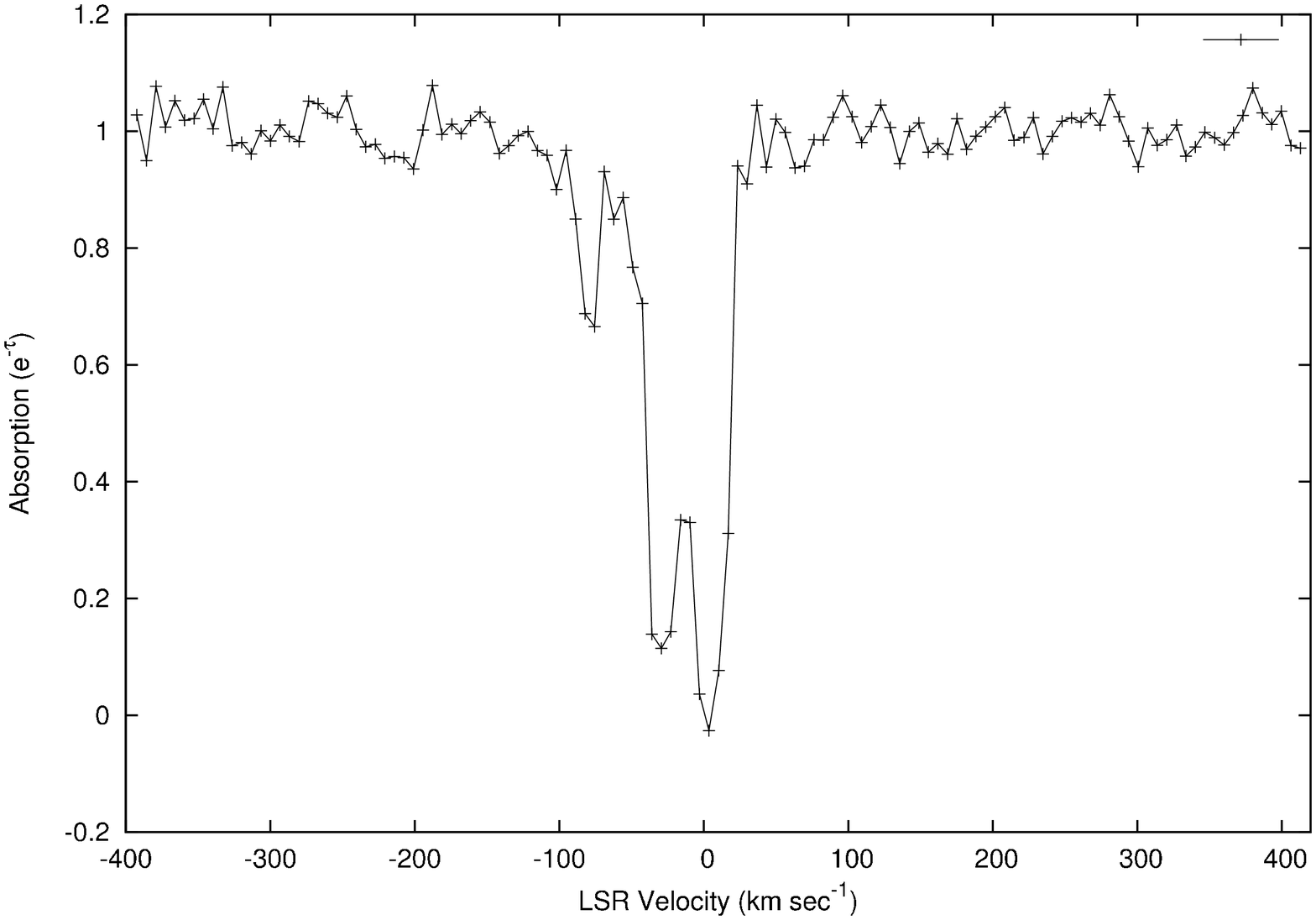}
  \caption{HI absorption spectrum towards the field HII region A (see text for
  details).} 
  \end{minipage}
  \end{figure}

\section{Discussions}

From the above, the morphology of the object G354.4+0.0 is found to be of a
partial shell. It shows polarized emission, and after correction of sky
background temperature, the spectrum of the shell is found to be steep that
holds good irrespective of whether it is at the center, in front or behind the
HII region.  Polarized emission along with inferred steep spectrum shows
synchrotron emission from the shell. 
Therefore, based on the shell type morphology, and synchrotron emission, the
shell type source G354.4+0.0 is classified as a newly discovered SNR. It is
also expected to emit in X-ray, but no counterpart is known in that band. The
surface brightness of the shell at 1 GHz is about 6 $\times 10^{-20}$
W~m$^{-2}$~Hz$^{-1}$~sr$^{-1}$, that compares well with the youngest SNR
G1.9+0.3 \citep{GREEN1984}. The orientation of magnetic fields in young SNRs is
typically radial \citep{MILNE1987}. Therefore, the polarization vectors
representing electric field direction in Fig.~1 should be cross-radial or
tangential to the shell in the absence of any Faraday rotation. 
We note the resolution of the NVSS (45$^{''}$) is not much smaller than the
size of the shell. This causes smearing of any change of the polarization
vectors with change in position on the shell.  From Fig.~1, we note the
polarization vectors in the northern part of the shell lies along the shell,
while only fraction of the polarized vectors in the southern part lies along
the shell.  Assuming magnetic field strength in the diffuse HII region to be of
typical ISM value of $\sim$10 $\mu$Gauss and the SNR at the center of the HII
region, given the electron density and the path length through the ionized gas,
any polarized emission from the SNR would undergo Faraday rotation by $\sim$60
radian at the observing wavelength of NVSS. Faraday rotation towards the
Northern part need to be 
$n\pi$ ($n$ $-$ integer) radian to explain the observation. It is possible that
the Faraday rotation towards the Southern part of the shell is slightly
different than the northern part of shell, and this causes a change in
direction of the observed polarization vectors from the tangential direction of
the shell.  Future polarization observations at three or more frequencies with
high angular resolution will be able to measure Faraday rotation accurately and
could show whether the shell lies inside or in front of the HII region.

\subsection{Evolutionary phase of the SNR and its age}

If the SNR is located outside the HII region, it is then expanding in typical
ISM. Given the assumed distance and the angular size of the remnant, it will be
in the free expansion stage.  Hence, the structure of the shock front and the
corresponding shock acceleration of energetic particles would not depend
significantly on the structure of the ISM in which it is expanding. Therefore,
the synchrotron emission from the shell should roughly be circularly symmetric.
However, the shell appears fragmented towards East, and only some parts of it
are seen in Fig.~1.  This is possible if the shell is no longer in the free
expansion stage.  Given the angular size and the distance to the SNR ($\sim$8
kpc), its linear diameter is about 3.7 pc. If it is within the dense HII
region, given its $n_e$, the SNR has swept up about 20~$M_{\odot}$ of
material from the surrounding ISM during its expansion. Kinetic energy of
ejecta from a supernova explosion lies in the range of $\sim 10^{51} - 10^{52}$
ergs \citep{WOOSLEY2006}.  Depending on the type of explosion, the ejecta mass
varies from $\sim 0.4~M_{\odot}$ (type Ia) \citep{STRITZINGER2006} to $\sim
20~M_{\odot}$ (core collapse) \citep{TADDIA2012}.  type of the progenitors and
the explosion, the
As the mass of the swept up material in this case is comparable to or more than
the typical ejected mass from supernova explosion, there will be appreciable
slowdown of its expansion. This phase of expansion is known as the adiabatic
phase \citep{WEILER1988}, and given the morphology discussed above, the SNR is
likely expanding in a denser than ISM environment. Assuming energy conservation
between the supernova ejecta and the swept up material and given the range of
initial ejecta mass, the age of the SNR is estimated from numerical
integration of $\frac{\tmop{dr}}{v ( r)}$ considering its slow down due to
expansion in the dense environment. The range of age possible for this remnant
is found to be $\sim$100$-$500 years. We note that the $T_e$ of 2300 K
determined in Sect.~3.2 is low for a typical HII region, that is typically
5000$-$10000 K.  However, WIM with temperatures of 3000 K has been reported
\citep{DOBLER2009} and unstable WIM can reach temperatures much lower
\citep{DONG2011}.  Moreover, averaging of flux densities within the angular
size of the source with likely spatial variation of electron density leads to
an underestimation of source temperature \citep{RUBIN1969}. Even if we assume
$T_e$ is a factor of three higher (7000 K), estimated value of $n_e$ being
mainly dependent on higher radio frequency emissions ($\gtrsim$1 GHz),
it will only increase by a factor of 1.2. This does not significantly change
the evolutionary phase of the SNR.

If a pulsar was created during the supernova explosion, given their typical
velocity of a few hundred km~s$^{-1}$ \citep{MOTCH2009}, it would be seen
within $\sim 10 ^{''}$ arc seconds from the center of the remnant.  However,
no pulsar is known within several arc-minutes of the object. This could
be caused by (i) a type-Ia explosion that does not leave behind any neutron
star, (ii) lack of beaming of any pulsar in the region towards us, or (iii)
lack of sensitive survey with suitable dedispersion of the pulses at the right
DM (the diffuse HII region itself would contribute a DM of about 400
cm$^{-3}$~pc).

\subsection{Surface brightness and expansion of G354.4+0.0 in the dense medium}

When an SNR expand inside a dense ISM, it is expected to be brighter. For
example, in M82, where emission measure of foreground gas towards the SNRs in
that galaxy is about four orders of magnitude higher than the medium
around G354.4+0.0 shell \citep{MCDONALD2002}, the typical surface brightness
of SNRs of the diameter of G354.4+0.0 are a few hundred times more
\citep{UROSEVIC2005}.  In our Galaxy, such a brightening has also been observed
for SNRs interacting with molecular clouds \citep{PAVLOVIC2013}. However,
particle density in the diffuse HII region is much less than that of typical
molecular clouds and that may explain why this SNR is not that bright.
Moreover, there is more than an order of magnitude spread of surface brightness
from what is predicted by $\Sigma -D$ relation \citep{PAVLOVIC2013}. The
surface brightness of G354.4+0.0 compares well with that of CTB 33
(G337.0$-$0.1) in our Galaxy expanding in a dense environment
\citep{PAVLOVIC2013}.

\section*{Acknowledgment}

We thank the staff of the GMRT that made the observations presented in this
paper possible. GMRT is run by the National Centre for Radio Astrophysics of
the Tata Institute of Fundamental Research. S.R. thanks Dave Green for useful
discussion. We thank the anonymous referee for help in improving the quality of
the paper. We also thank Poonam Chandra and Arunima Banerjee for going through
the manuscript.  Work of SP is supported by a grant of MOES.

\end{document}